\documentclass[lettersize,journal]{IEEEtran}
\usepackage{amsmath,amssymb,amsfonts}
\usepackage{algorithmic}
\usepackage{algorithm}
\usepackage{array}
\usepackage[caption=false,font=normalsize,labelfont=sf,textfont=sf]{subfig}
\usepackage{textcomp}
\usepackage{stfloats}
\usepackage{url}
\usepackage{verbatim}
\usepackage{graphicx}
\usepackage{cite}
\usepackage{xcolor}
\usepackage{dsfont}
\usepackage{amsthm}
\usepackage{xcolor}
\usepackage{tabularx}
\usepackage[export]{adjustbox}
\usepackage{subcaption}
\usepackage{soul}

\newtheorem{lem}{Lemma}
\theoremstyle{definition}
\newtheorem{definition}{Definition}[section]
\newcommand{\Hyp}[1]{\ensuremath{\mathbb{H}_{#1}}}

\newcommand{\testH}{\ensuremath{\underset{\Hyp{0}}{\overset{\Hyp{1}}{\gtrless}}}}

\begin{document}

\title{Neyman Pearson Detector for Multiple Ambient Backscatter
Zero-Energy-Devices Beacons using Near-Perfect Code}

\author{ Shanglin Yang*\textsuperscript{o}, Jean-Marie Gorce\textsuperscript{o}, Muhammad Jehangir Khan\textsuperscript{o}, Dinh-Thuy Phan-Huy*, Guillaume Villemaud\textsuperscript{o} \\

*Orange Innovation/Networks, Châtillon, France \{name.surname\}@orange.com\\

\textsuperscript{o} INSA Lyon, Inria, CITI, EA3720, Villeurbanne, France \{name.surname\}@insa-lyon.fr
}

\markboth{IEEE Journal of Radio Frequency Identification}%
{Shell \MakeLowercase{\textit{et al.}}: A Sample Article Using IEEEtran.cls for IEEE Journals}


\maketitle


\begin{abstract}
Recently, a novel ultra-low power indoor localization system based on Zero-Energy-Devices (ZEDs) has shown promising results in ambient backscatter communication. In this paper, we study detection of multiple coexisting Zero-Energy Devices (ZEDs) in ambient backscatter systems under interference and synchronization uncertainty. Building on a Neyman–Pearson (NP) formulation previously applied to single-tag detection, we introduce a detector tailored to multi-tag scenarios. The core idea is to use a Near-Perfect Code (NPC) as the synchronization sequence, which substantially improves the peak-to-sidelobe (PSL) ratio and thus separability among concurrent tags. The proposed scheme replaces dual band-pass filtering with dual correlators, enabling an explicit Bayesian detector and tight control of the false-alarm rate; we further incorporate a contrast metric and multi-frequency combining to reveal secondary tags. Experiments on the CorteXlab testbed (part of the SLICES-EU infrastructure) confirm robustness at low SNR, with observed PSL improvements from \(\approx 11\) dB to \(>21\) dB. These results advance scalable, reliable ambient backscatter localization in practical multi-tag environments.
\end{abstract}


\section{Introduction}
\IEEEPARstart{A}{mbient} backscatter (AmB) \cite{AmB} has emerged as a promising technology for the Internet of Things (IoT), gaining increasing interest from both academia and industry. It enables IoT devices to transmit data by reflecting existing ambient radio-frequency signals from ambient wireless networks, without generating their own carrier wave. It advantageously enables IoT detection by any smartphone connected to an ambient wireless network, rather than requiring deploying specialized readers like RFID readers. Additionally, it achieves ultra low power consumption at the transmitter side, making it highly suitable for energy-constrained IoT devices. Among the candidate technologies for the sixth generation (6G) of wireless networks, as outlined in the Hexa-X I European Flagship Project, the groundbreaking concept of Zero-Energy-Device (ZED) has been introduced \cite{6GHexa-X,InnovHexa-X} to provide IoT services in a sustainable manner. A ZED is an energy-autonomous device that self-powers by harvesting energy from ambient sources like radio-frequency (RF), light, heat or vibration. Several flavours of ZED, called "Ambient Internet of Things" (A-IoT) devices, also relying on backscattering, are currently being discussed for inclusion in the 3rd Generation Partnership Project (3GPP) for 5G-advanced standard,  \cite{3gpp_tr_38_848}. One particular flavour of ZED, called crowd-detectable ZED (CD-ZED), leverages light-energy harvesting to power itself, and exploits ambient backscatter communication to reuse the ambient cellular network infrastructure, smartphones and spectrum, with minimum additional costs. CD-ZED has been proposed to serve as zero-energy wireless beacons for indoor localization services or tags for asset tracking \cite{InnovHexa-X,AmB-ZED-DT}.  Several studies \cite{AmB-ZED-DT, AmB-FSK-Alto, AmB-DemoLTE-Alto, ZED-6G-Papis}, have presented proof-of-concept (PoCs) and live-demonstrations of CD-ZEDs. Also, the detection of CD-ZEDs (based on a Neyman Pearson approach \cite{AmB-detection-zargari2024}) has been modeled and the coverage of CD-ZEDs has been simulated \cite{ZED-Loc-SY}. All these previous studies where limited to single CD-ZED detection. Existing previous studies on multiple ZED detection, such as independent component analysis based blind signal separation \cite{Multi-tag-CDMA} and Non-Coherent Parallel Detection \cite{Multi-tag-NCPD} have focused on the challenge of multiple ZED identification, once all ZEDs are perfectly synchronized. Hence, up to now, to our best knowledge,  no solution has been proposed to solve the problem of synchronization in a multiple CD-ZED scenario.

In this paper, for the first time, we propose a novel method to solve the problem  of synchronization in a multiple CD-ZED scenario. For that purpose, we propose to use the near perfect code (NPC) as the synchronization sequence of CD-ZEDs and we propose a novel detector for multiple-ZEDs , building on the system initiated in \cite{yang2025neymanpearsondetectorambientbackscatter}. Therefore, as in \cite{yang2025neymanpearsondetectorambientbackscatter}, the proposed detector is optimized using the Neyman-Pearson theory \cite{neyman1933}. Compared to \cite{yang2025neymanpearsondetectorambientbackscatter}, while we build up upon the original CD-ZED technology and coded sequences, we substitute the received dual frequency band-pass filters with dual correlators. This modification enables a precise determination of the optimal Bayesian detector and facilitates control over the false alarm probability, which was not addressed in the initial design. Furthermore, the framework presented here leverages the Neyman-Pearson formalism. This framework is key for CD-ZED coverage planning applications, as it allow for the detection performance to be simulated and evaluated based upon factors such as synchronization sequence length and coverage requirements. Unlike previous studies \cite{AmB-ZED-DT, AmB-FSK-Alto, AmB-DemoLTE-Alto, ZED-6G-Papis}, this paper is the first to offer a formal evaluation of multiple CD-ZED performance. Our approach is validated thanks to experimental measurements using the CorteXlab testbed \cite{massouri2014cortexlab}.
For convenience, in the rest of this paper, we will call the CD-ZED simply ZED. 
The paper is organized as follows: Section II presents our novel detection model for ZEDs, Section III presents our performance evaluation based on experiments, and Section IV concludes this paper.

\begin{figure}[htbp]
\centerline{\includegraphics[scale=0.6]{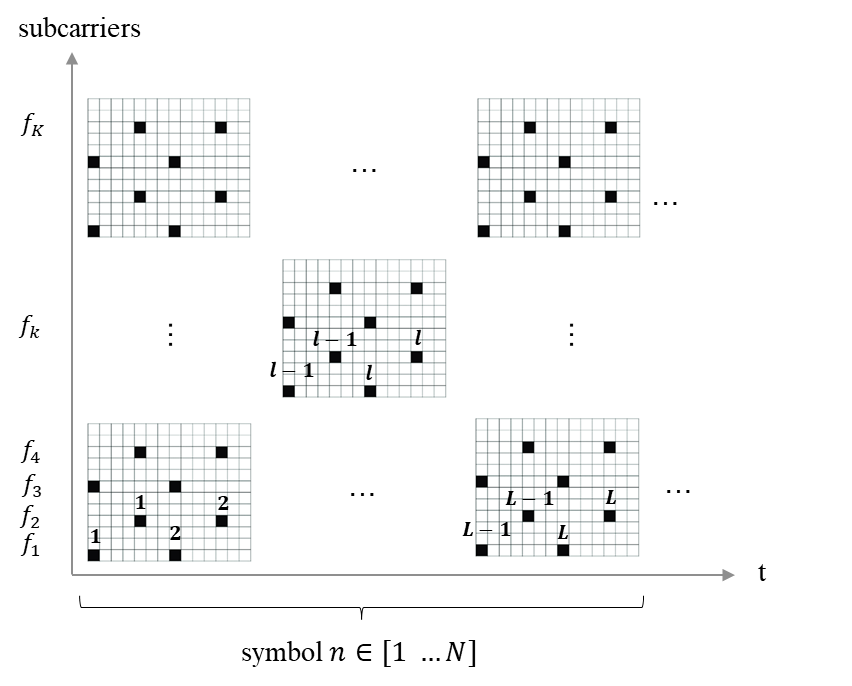}}
\caption{Structure of LTE reference signal pilots in the OFDM time-frequency grid, showing the RS placement used for ZED synchronization and signal reflection.}
\label{fig:pilots}
\end{figure}


\section{Detection Model for ZEDs}\label{sec:zedDes}


This section presents the proposed detection framework for identifying ZEDs in ambient backscatter systems. The system leverages synchronization sequences based on NPC and operates without requiring coordination between tags. Our approach is structured to address both single-tag and multi-tag scenarios in asynchronous and low-SNR environments. The detection pipeline consists of a ZED transmitter model, signal reception and correlation, contrast computation, and a threshold-based decision rule derived from the Neyman-Pearson criterion. The framework is further extended to enable reliable detection of secondary ZEDs by exploiting the autocorrelation properties of NPC. Each component is detailed in the following subsections.


Section~\ref{subs:trasnmistter} introduces the proposed ZED transmitter system and its signal generation mechanism.  
Section~\ref{subs:Det_ccpt} details the construction of an optimal detector.
Section~\ref{subs: NP} details the Neyman-Pearson hypothesis testing procedure used to establish a detection threshold under a fixed false alarm constraint.
Section~\ref{subs:Det_multitag} extends the detection strategy to the multi-tag scenario by introducing a contrast-based metric to identify secondary ZEDs in the presence of multiple active tags.

\subsection{ZED equivalent transmitter}\label{subs:trasnmistter}
\begin{figure}[htbp]
\centerline{\includegraphics[scale=0.42]{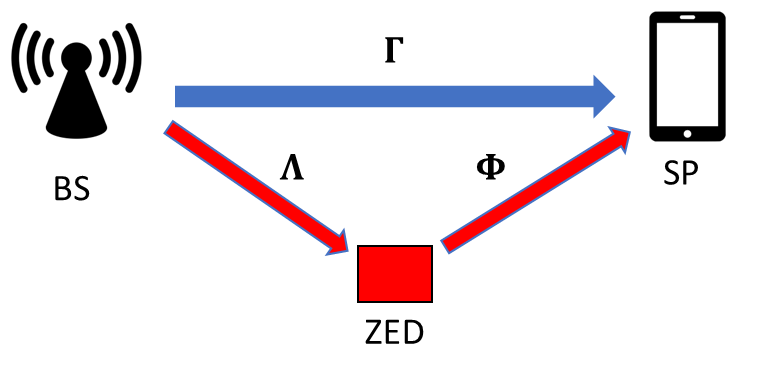}}
\caption{Single-ZED communication channel model}
\label{fig:ZED_model1}
\end{figure}

In this subsection, we briefly summarize the single-tag detection model. The system employs ambient backscatter technology, where a ZED modulates and backscatters ambient signals from cellular base stations to communicate its unique identifier. 
Specifically, the ZED operates in two impedance states, often referred to as 'reflective' and 'transparent' modes. However, in both states, the tag continues to backscatter the ambient signal. The difference lies in the slight variation in the amplitude and phase of the backscattered signal, caused by the load modulation, which results in a detectable change in the channel response at the smartphone (SP) receiver, as illustrated in Fig.\ref{fig:ZED_model1} .

The interaction between a ZED and a SP is enabled by leveraging ambient downlink signals from LTE base stations. 

The LTE reference signals (RS) mapped on specific resource elements of the OFDM frame structure are illustrated in Fig.~\ref{fig:pilots}. Let $m:[1,K]\times [1,L] \rightarrow t\in \mathbb{R^+}$ noted $m(k,l)$ be the mapping function between carrier frequency index $k\in(1,K)$ and RS index $l\in(1,L)$ and time $t\in \mathbb{R^+}$, this function is illustrated in Fig. 1. In accordance with 4G network standards, these RS are distributed over \(K = 4N_{RB}\) subcarriers. Each Transmission Time Interval (TTI) has 14 OFDM symbols, with a total duration of \(\text{TTI} = 14T^{\text{ofdm}}\) \cite{3gpp_ts_36_211}., where \(T^{\text{ofdm}}\)  is the OFDM symbol duration. For analytical simplicity, the RS are modeled as having uniform amplitude \(\sqrt{P_u}\), where \(P_u\) denotes the power allocated to the pilot signals.

\begin{figure}[htbp]
\centerline{\includegraphics[scale=0.7]{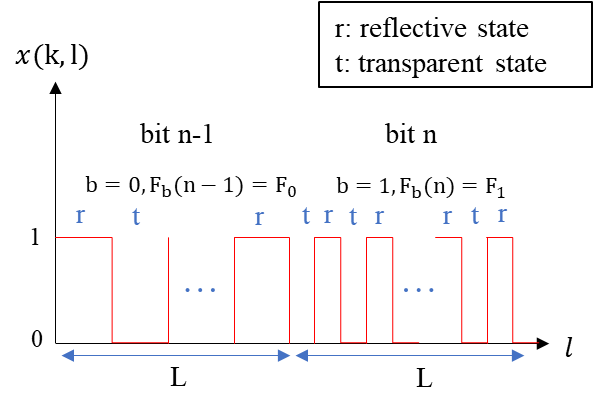}}
\caption{Signal generated by the ZED for two successive bits with $b_{n-1}=0$, $b_{n}=1$.}
\label{fig:tag_state}
\end{figure}

The transmission sequence of a ZED consists of two parts: a synchronization sequence and an identification sequence. In this paper, we focus exclusively on the synchronization sequence. The synchronization sequence is generated using frequency-shift keying (FSK) modulation applied to orthogonal Hadamard sequences. 
To transmit a bit 0 (or 1), the ZED generates a periodic signal at frequencies $F_0$ (or $F_1$), as illustrated in Fig.\ref{fig:tag_state}. The FSK symbol duration is chosen so that the induced artificial doppler spread is larger than natural doppler spread for pedestrian users, and small enough to be tracked by the channel estimator \cite{AmB-FSK-Alto}. Each transmitted bit duration is defined as \(T_b = \frac{L}{2}T_{\text{TTI}}\), with chip duration \(T_c\) and number of chips \(N_c\), where \(T_b = N_c T_c\). As we already introduced in \cite{yang2025neymanpearsondetectorambientbackscatter}, The synchronization signal for a pseudo-random bit sequence \(b = [b_1, ..., b_{N_b}]\) is expressed as:
\begin{equation}
\label{eq:refseq}
    x(t)=\sum_{n=0}^{N_b-1}b_n x_1(t-nT^b)+\sum_{n=0}^{N_b-1}(1-b_n)x_0(t-nT^b),
\end{equation}
and is depicted in Fig.\ref{fig:tag_state}.
We note, for mathematical convenience, $x(k,l)=x(t=m(k,l))$ corresponding to the sampled version of this sequence, observed by the receiver at the RS positions. 

As we already introduced in \cite{yang2025neymanpearsondetectorambientbackscatter}, the received signal $y(m(k,l))$ noted as $y(k,l)$ at sub-carrier \(k\) and RS symbol \(l\) is:
\begin{equation}
y(k,l)=e^{j\phi(k,l)}\sqrt{P_u}\left( \mathbf{\Gamma} (k)+\mathbf{\Lambda}(k)\mathbf{\Phi}(k)x(k,l)\right)+\alpha(k,l)
\end{equation}
where \(\Gamma(k)\) and \(\Lambda(k)\Phi(k)\) represent direct and reflected channel responses respectively. $\alpha(k,l) \sim \mathcal{CN}(0, \sigma^2)$ denotes a complex Gaussian noise component with zero mean and variance $\sigma^2$, while $\phi(k,t)$ represents phase fluctuations occurring across successive TTIs, primarily caused by imperfections in our receiver's local oscillator. 
Indeed, our receiver is implemented using an SDR platform with hardware impairments. In a commercial LTE smartphone, there would not be such impairments and therefore a coherent detector could be used instead. Since this phase distortion is unknown and cannot be compensated, a non-coherent reception strategy is required. 
For now, the firmware of the smartphone would need to be changed. In addition to the existing firmware for coherent detection of data, a non-coherent detection branch should be added for zed detection.

To enable the detection of multiple ZEDs in the environment, the peak-to-side-lobe (PSL) ratio is a critical metric, as side-lobes can interfere with the detection of concurrent tags. In our previous work, we utilized a Barker code, which offered a PSL gain of approximately 11 dB \cite{yang2025neymanpearsondetectorambientbackscatter}. To further suppress interference and enhance detection accuracy, we propose using a NPC that provides a substantially improved PSL ratio. For example, a 25-bit NPC achieves a PSL gain of 21.93 dB, and is defined as \cite{meikle2001modern}:
\[
\text{NPC} = [0 \ 1 \ 1 \ 0 \ 1 \ 1 \ 0 \ 1 \ 0 \ 1 \ 0 \ 1 \ 1 \ 1 \ 1 \ 1 \ 1 \ 0 \ 0 \ 0 \ 1 \ 1 \ 0 \ 0 \ 0]
\]
This significant improvement in autocorrelation characteristics enables more reliable detection in multi-tag scenarios. All simulation results presented in Section \ref{sec:Res} are obtained with this NPC sequence.

\subsection{Detection System concept}\label{subs:Det_ccpt}

\begin{figure*}[t]
\centering
\includegraphics[scale=0.5]{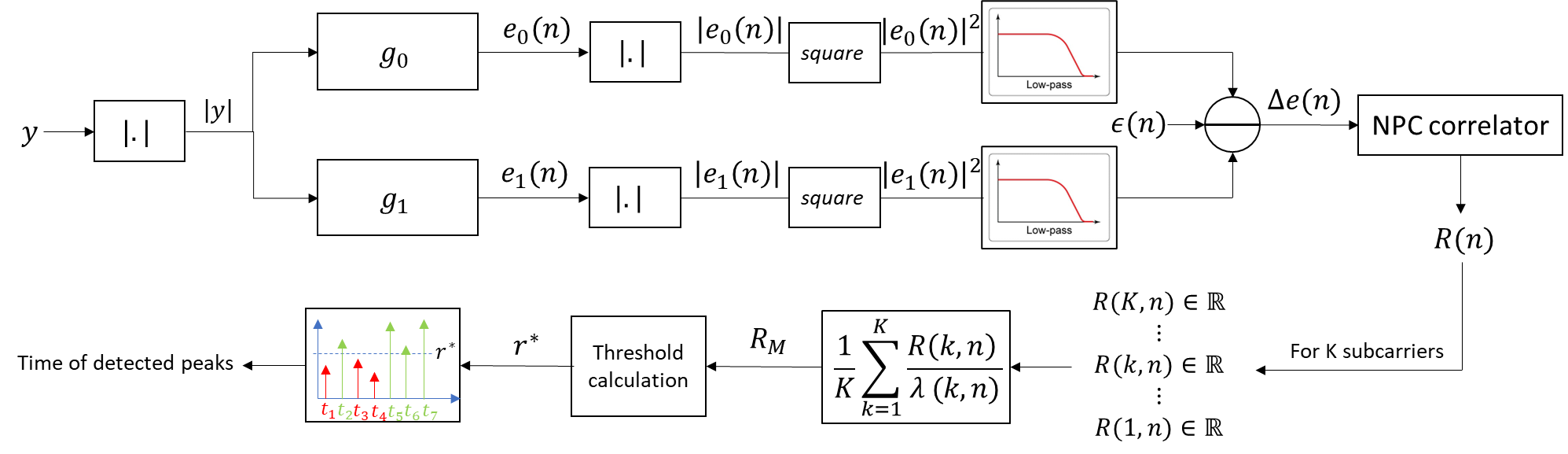}
\caption{Detection system concept. The architecture includes dual correlators, low-pass filtering, and a contrast computation stage based on NPC sequences, enabling robust detection of ambient backscatter signals in asynchronous multi-tag environments.}
\label{fig:Ccpt_sys}
\end{figure*}

To improve robustness and adaptability of ZED detection in asynchronous and multi-signal environments, we propose an enhanced architecture composed of three main modules: two adaptive correlators, a low-pass filter, and an NPC sequence combiner. This modular design allows the receiver to compensate for irregular sampling, time misalignments, and low signal levels through a structured processing pipeline, as illustrated in Fig.~\ref{fig:Ccpt_sys}.

For each OFDM symbol with index \(n\) and transmitted at time $t_n$, the received signals are processed by correlators matched to the transmitted sequences \(x_0(t)\) and \(x_1(t)\). The two correlators process incoming signals to extract the relevant reflections caused by ZED activity. For each OFDM symbol index \(n\) and transmitted at time $t_n$, the correlator outputs $e_0(n)$ and $e_1(n)$ are computed by aggregating samples within the bit interval $[t_n,t_n+T_b]$, and normalizing based on the number of samples in each mode. The correlator outputs are:

\begin{equation}
    \begin{split}
    \label{eq:e0e1}
    e_0(n)&=\sum_{\tiny\begin{matrix}l; t(l)\in \\ [t_n,t_n+T^b]\end{matrix}}  \frac{y(l)}{a_{0}(n)}\cdot \mathds{1}_{x_0(t(l))} - \frac{y(l)}{b_{0}(n)}\cdot \mathds{1}_{\bar{x_0}(t(l))} \\
    e_1(n)&=\sum_{\tiny\begin{matrix}l; t(l)\in \\ [t_n,t_n+T^b]\end{matrix}}   \frac{y(l)}{a_{1}(n)}\cdot \mathds{1}_{x_1(t(l))} - \frac{y(l)}{b_{1}(n)}\cdot \mathds{1}_{\bar{x_1}(t(l))} 
    \end{split}    
\end{equation}
Here, $\mathds{1}{c}$ denotes the indicator function, equal to 1 if condition $c$ is true and 0 otherwise. The terms $a_i(n)=\sum_{l; t(l)\in [t_n,t_n+T^b]}  \mathds{1}_{(g_i(t(l))=1}$ and $b_i(n)=\sum_{l; t(l)\in [t_n,t_n+T^b]}  \mathds{1}_{(g_i(t(l))=-1}$. $a_i(n)$ and $b_i(n)$ respectively count the number of samples under transparent and reflective modes, verifying $a_i(n) + b_i(n) = L$. They normalize the contributions in \eqref{eq:e0e1}, ensuring the cancellation of the direct path component $|\Gamma|$ even under irregular sampling, and allow resampling at the desired rate. These pseudo-filters can be implemented efficiently with a computational cost similar to that of conventional filters.

To address temporal uncertainty due to clock drift or asynchronous transmissions, the correlator outputs are passed through a low-pass filter with a cutoff frequency tuned to preserve significant bit-level transitions while suppressing noise and small misalignment. This enhances peak stability across detection intervals.

When a ZED is active, it reflects either the sequence $x_0(t)$ or $x_1(t)$, depending on the transmitted bit $i \in \{0,1\}$. Let $j = 1 - i$ be the complementary bit. If the correlators are matched to $x_i(t)$ and $x_j(t)$ respectively, and assuming perfect 
reflection and transparent, thanks to our previous work in \cite{yang2025neymanpearsondetectorambientbackscatter}, the correlator outputs can be approximated as:
\begin{equation}
e_i(n) = \sqrt{P_u} \gamma + \alpha_i(n), \quad e_j(n) = \alpha_j(n),
\end{equation}
where $\gamma$ represents the channel coefficient associated with the ZED, and $\alpha_i(n)$ and $\alpha_j(n)$ are zero-mean Gaussian noise terms. Their variances depend on the number of samples taken in transparent and reflective states:
\begin{equation}
\alpha_i(n) \sim \mathcal{N}\left(0, \frac{\sigma^2}{2} \left( \frac{1}{a_i(n)} + \frac{1}{b_i(n)} \right) \right),
\end{equation}
where $a_i(n)$ and $b_i(n)$ are the counts of samples under transparent and reflective modes during bit interval $n$. These counts are used to normalize the correlator output, ensuring cancellation of the direct path component even under irregular sampling. When sampling is regular and balanced, i.e., $a_i(n) = b_i(n)$, the noise variance is minimized \cite{yang2025neymanpearsondetectorambientbackscatter}.

Given these two outputs, we aim to estimate the strength of the backscattered signal. An unbiased estimator of the ZED path power $\eta^2 $ can be defined as \cite{yang2025neymanpearsondetectorambientbackscatter}:
\begin{equation}
\tilde{\eta}(n) = |e_i(n)|^2 - |e_j(n)|^2 - \epsilon(n),
\end{equation}
where the correction term $\epsilon(n)$ accounts for sampling imbalance:
\begin{equation}
\epsilon(n) = \sigma^2 \left( \frac{1}{a_i(n)} + \frac{1}{b_i(n)} - \frac{1}{a_j(n)} - \frac{1}{b_j(n)} \right).
\end{equation}
In the case of ideal sampling conditions with $a_i = b_i = a_j = b_j = L/2$, the correction $\epsilon(n)$ cancels out.

This estimator corresponds to the maximum likelihood (ML) estimation and is unbiased according to \cite{talukdar1991estimation}.

Subsequently, a contrast metric \(R(n)\) is computed by correlating the filtered outputs with the known NPC sequence:

\begin{equation}\label{eq:Rn}
\begin{split}
    R(n) &=  \frac{1}{N_b} \sum_{m=0}^{N_b-1} (2b_m - 1) \cdot (e_{b_m}(n + m\delta_t)^2\\
    & - e_{1 - b_m}(n + m\delta_t)^2-\epsilon(n + m\delta_t)) 
\end{split}
\end{equation}
with $\delta t=\frac{T^b}{T^{ofdm}}$.

In synchronized cases, invoking the central limit theorem, when $N_b$ is large enough, \(R(n)\) converges to the reflected path energy \(\eta^2\) when a ZED is working, and follows a Gaussian distribution with zero mean when there is no ZED. This contrast serves as the input to a Neyman-Pearson test that ensures reliable detection performance with controllable false alarm probability.

Although the Central Limit Theorem is asymptotic, we verified empirically that for $N_b$=25 (used in our experiments), the contrast metric $R(n)$ when there is no ZED closely follows a Gaussian distribution. Simulated false alarm rates match theoretical predictions derived from Eq. (11), confirming the validity of the Gaussian approximation in practical settings.

To account for the frequency-selective nature of the wireless channel, we extend the model to operate across multiple subcarriers. Let \(R(n, k)\) denote the contrast output on subcarrier \(k\). A multi-frequency combiner aggregates these outputs to improve robustness:
\begin{equation}
    R_M(n) = \frac{1}{K} \sum_{k=1}^{K} \frac{R(n, k)}{\lambda(n,k)},
\end{equation}
where \(\lambda(k,n)=\frac{1}{N_b}\left(\frac{1}{a_i(n,k)}+\frac{1}{b_i(n,k)} + \frac{1}{a_j(n,k)} + \frac{1}{b_j(n,k)}\right)\geq \frac{4}{L N_b}\) .

\begin{figure}[htbp]
\centerline{\includegraphics[scale=0.42]{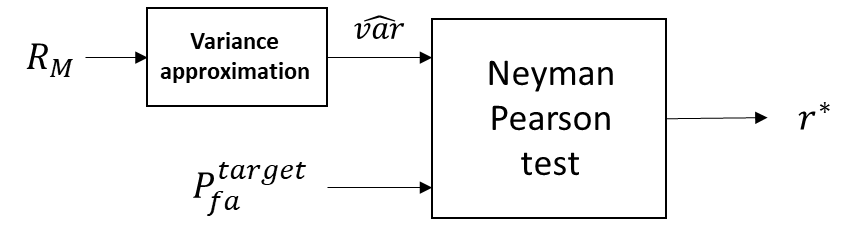}}
\caption{Calculation of threshold using Neyman Pearson test}
\label{fig:Threshold_calculation}
\end{figure}

This modular structure enhances detection reliability in practical environments by jointly exploiting temporal and frequency diversity. The final detection decision is performed by comparing \(R_M(n)\) with a threshold derived in the next subsection using the Neyman-Pearson test, as expressed in Fig. \ref{fig:Threshold_calculation}.

The hypothesis test corresponding to a given sequence is:
\begin{itemize}
    \item \Hyp{0}: The searched sequence is not present, and the contrast is just random due to the noise. $R_M(n)$ is normally distributed.
    \item $\Hyp{1|\eta^2}$: The ZED is present, with a given ZED path $\eta^2$.
\end{itemize}

To calculate the optimal threshold, we use the following lemma proposed in \cite{yang2025neymanpearsondetectorambientbackscatter}.
\begin{lem}\label{lem:Htest}
Given the hypothesis test above, any optimal decision in the Bayesian sense is given by 
\begin{equation}
R_M \testH r^*,
\end{equation}
where $R_M=\frac{1}{K}\sum^{K}_{k=1} \frac{R(n,k)}{\lambda(n,k)}$ is the average of the likelihood factor over all the subcarriers.
\ $r^*$ is the decision threshold. 
\end{lem}

While the test statistic $R_M(n)$ is derived from the single-tag Neyman-Pearson framework, it serves as a practical approximation for multi-tag detection by treating non-dominant tags as noise. A full derivation of the optimal likelihood ratio test for multiple coexisting, asynchronous ZEDs is non-trivial and left for future work. Nonetheless, $R_M(n)$ provides robust performance in practice, as confirmed by experiments.

\subsection{Neyman-Pearson Test} \label{subs: NP}

When no ZED is active in the environment, we can calculate the optimal detection threshold \( r^* \) using the estimated noise variance \( \widehat{var} \) and a predefined false alarm probability. Normally, determining \( r^* \) would require prior knowledge such as the costs of false alarms and missed detections, the probabilities of hypotheses \(\mathcal{H}_0\) and \(\mathcal{H}_1\), and the value of the unknown signal strength \(\eta^2\) from the ZED.

To avoid relying on these unknowns, we adopt the Neyman-Pearson (NP) approach \cite{poor2013}, which allows us to set a target false alarm probability \( p^{target}_{fa} \) from the application’s requirements—for example, how often we can tolerate detecting a ZED when none is actually present. With this probability, we can compute the threshold \( r^* \), and then estimate the corresponding detection probability \( p_D(\eta^2) \), which depends on the strength of the ZED signal.

Although the NP framework has been applied to ambient backscatter in \cite{AmB-detection-zargari2024}, their work uses different signal and noise models and does not consider the contrast-based detection we propose here.

\begin{lem}
According to our previous work in  \cite{yang2025neymanpearsondetectorambientbackscatter}, the false alarm probability \( p_{fa} \) associated with the detector described in Lemma~\ref{lem:Htest} is given by:
\begin{equation}
\label{eq:pfa}
    p_{fa} = \mathbb{P}_{R_M|\mathcal{H}_0}[r > r^*] = Q\left(\frac{r^*}{\sqrt{\widehat{var}}}\right)
\end{equation}

The threshold \( r^* \) corresponding to a target false alarm probability \( p^*_{fa} \) is:
\begin{equation}
\label{eq:seuil}
    r^* = \sqrt{\widehat{var}} \times Q^{-1}(p^*_{fa})
\end{equation}

The detection probability as a function of the signal strength \( \eta^2 \) is then:
\begin{equation}
\label{eq:pd}
    p_D(\eta^2) = \mathbb{P}_{R_M|\mathcal{H}_1}[r > r^*] = Q\left(\frac{r^* - \eta^2}{\sqrt{\widehat{var}}}\right)
\end{equation}
\end{lem}

\subsection{Detection for multiple ZEDs}\label{subs:Det_multitag}

\begin{figure}[htbp]
\centerline{\includegraphics[scale=0.42]{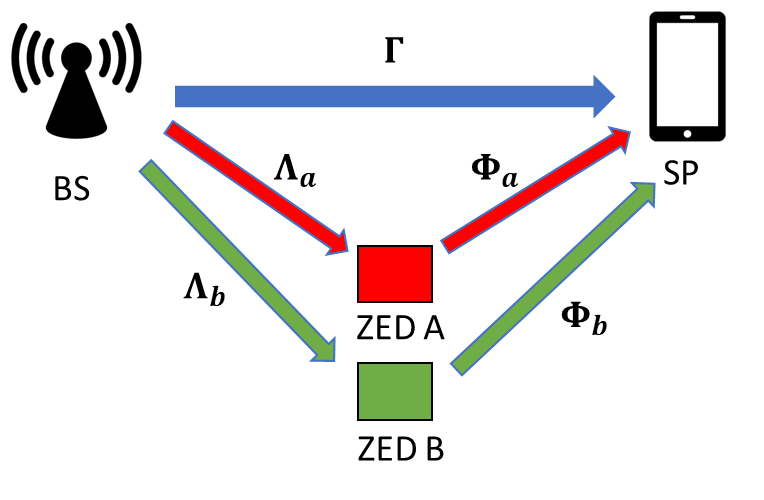}}
\caption{Multiple-ZED based communication channel model}
\label{fig:ZED_model2}
\end{figure}

In this subsection, we consider the presence of two ZEDs in the environment, referred to as ZED A and ZED B, as illustrated in Fig.~\ref{fig:multitag}. Both ZEDs employ the same synchronization sequence for transmission. To enable distinguishable detection, each ZED is assigned a different waiting time, denoted by \( T_{\text{wait}}^A \) and \( T_{\text{wait}}^B \), respectively. We apply different sequence time for these two ZEDs in the experimental setup to permit the obervation of second peaks at different distances from the main peak, within a unique experiment. Despite that such timing differentiation is not strictly required in real applications, this may be an interesting way to increase multi-tag detection, but at the price of an increased processing time. Note that during its waiting period, the corresponding ZED remains in a transparent state and does not reflect any signal. The complete transmission cycle for each tag, including both the waiting and active periods, is therefore defined as:
\begin{equation}
    T_A=T_{seq}+T_{wait}^{A},T_B=T_{seq}+T_{wait}^{A}
\end{equation}
where $T_{seq}$ is the duration of the NPC sequence.

\begin{align}
y(k, l) =\ & e^{j\phi(k, l)} \sqrt{P_u} \Big[ \Gamma(k) 
+ \mathds{1}_{\text{A}}(l)\, \Lambda(k)\Phi_A(k)\, x(k, l - l_{\text{offset}}^A) \nonumber\\
& + \mathds{1}_{\text{B}}(l)\, \Lambda(k)\Phi_B(k)\, x(k, l - l_{\text{offset}}^B) \Big]
+ \alpha(k, l)
\label{eq:multiZED}
\end{align}
where  $\mathds{1}_{\text{A}}(l) = 1$ if $t(l) \bmod T_A \in [T_{\text{wait}}^A,\ T_A)$ , 0 otherwise; $\mathds{1}_{\text{B}}(l) = 1$ if $t(l) \bmod T_B \in [T_{\text{wait}}^B,\ T_B)$ , 0 otherwise; $l_{\text{offset}}^A = \left\lfloor \frac{t(l) - T_{\text{wait}}^A}{T^{\text{ofdm}}} \right\rfloor \bmod N_{\text{seq}}$, only valid if tag A is active at time $l$; $l_{\text{offset}}^B$ is defined similarly for tag B; $N_{\text{seq}}$ is the number of OFDM symbols spanned by one sequence

In this work, the ZED transmission cycles $T_A$ and $T_B$ are predetermined to avoid overlap and enable clear separation of peaks. This setup reflects a practical deployment scenario where ZED coverage is spatially limited (typically within a few meters\cite{ZED-Loc-SY}), allowing designers to control overlap through careful placement and configuration. Such planning is feasible in many real-world use cases such as indoor localization or asset tracking.

\begin{figure}[htbp]
\centerline{\includegraphics[scale=0.4]{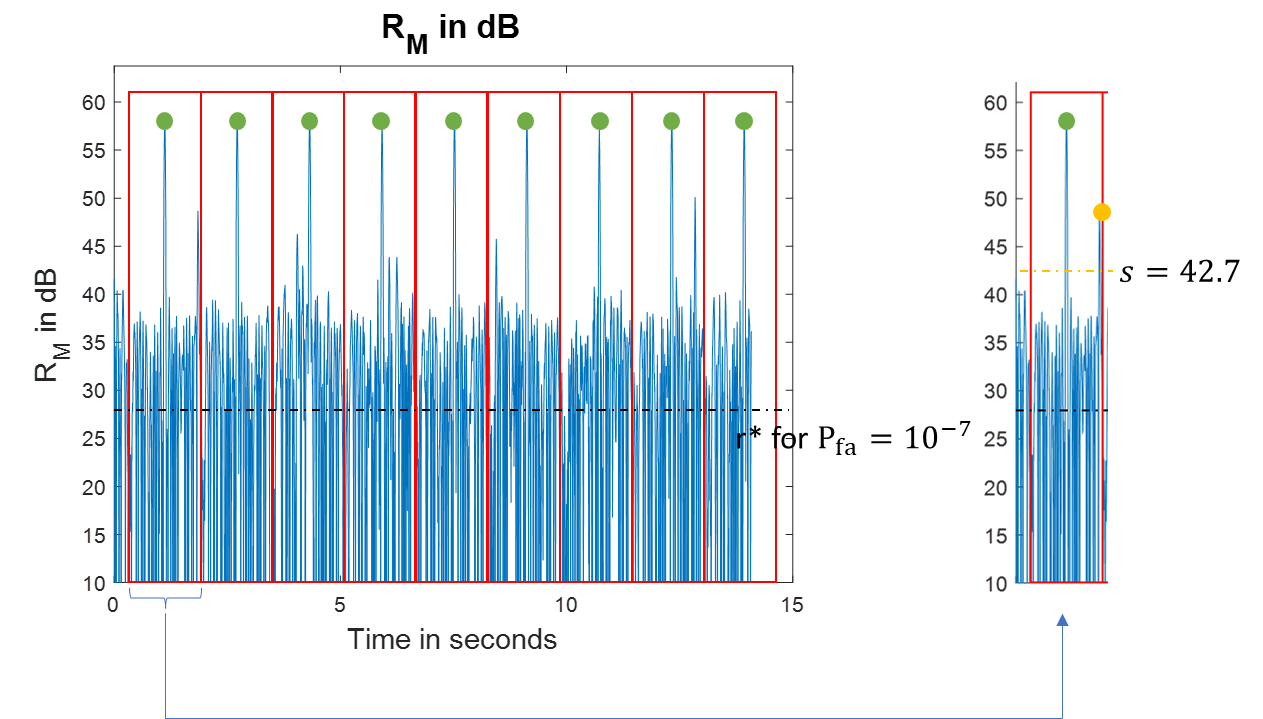}}
\caption{Example for the detection of the second peak in a 15 seconds window}
\label{fig:sec_det}
\end{figure}

To detect two coexisting ZEDs, 
by assuming the ZED period $T_A$ and $T_B$ are already known by the SP, we begin by observing the received signal over one period of duration $T_{\text{obs}} = \min(T_A, T_B)$, where $T_A$ and $T_B$ are the full cycles (waiting + sequence) of ZED A and ZED B, respectively. Within each observation window of duration $T_{obs}$, we first identify the strongest peak in the contrast output $R(n)$, which is assumed to correspond to the dominant ZED. Next, we fix the strongest peak in the middle and search for a secondary peak in the vicinity of the first one, within a window length equal to $T_{obs}$. This process is illustrated in Fig.\ref{fig:sec_det}.

Assume that ZED A produces the strongest signal. As shown in Fig.~\ref{fig:Ccpt_sys}, and using the detection threshold $r^*$ derived in the previous section, ZED A can be reliably detected even if the signal from ZED B is treated as additional noise or interference. The challenge, then, lies in identifying ZED B, whose peak may be masked by the side-lobes generated by ZED A’s signal.

To address this, we recall from Section~\ref{subs:trasnmistter} that the NPC provides a PSL gain of up to 21.93 dB. This property allows us to define an additional detection threshold that helps distinguish genuine peaks from side-lobes.

\begin{definition}[Secondary ZED Detection Threshold]
Let $P^{\text{dB}}_0$ denote the power (in dB) of the strongest detected peak, assumed to belong to ZED A. Then, the threshold for detecting the second ZED is defined as:
\begin{equation}
    s = P^{\text{dB}}_0 - G_{\text{PSL}} + M
\end{equation}
where $G_{\text{PSL}} = 21.93$ dB is the PSL gain of the NPC sequence, and $M$ is a margin factor accounting for noise and interference variability.
\end{definition}

A secondary ZED is considered detected if a second peak is found such that its power exceeds both the Neyman-Pearson threshold $r^*$ and the side-lobe suppression threshold $s$. This is illustrated in Fig. \ref{fig:sec_det}.

\section{Performance Evaluation by Experiment}\label{sec:Res}
To assess the effectiveness of the proposed multi-ZED detection model, we conducted experiments on the CorteXlab testbed \cite{massouri2014cortexlab}. This testbed supports up to 40 software-defined radio nodes and enables controlled and reproducible wireless experimentation. In this study, we employed a two-node configuration: one node emulating the base station transmitter and the other serving as the receiver that captures and processes the ambient backscatter signals.
\subsection{Experimental setups}
The experiments were performed in a controlled indoor environment (Figure \ref{fig:scn0}, \ref{fig:monotag} and \ref{fig:multitag}) to evaluate the system’s capability in detecting multiple coexisting ZEDs. The detection performance was analyzed by measuring key metrics such as peak-to-lobe ratio and detection probability under varying noise conditions.

We used LTE-based ambient signals where each frame contains 14 OFDM symbols, including 2 reference signals utilized by the ZEDs. The experimental setup includes the following parameters: bandwidth $B = 2.5$ MHz, FFT size of 128, and an 8-subcarrier cyclic prefix (CP). Each OFDM symbol, including the CP, spans a duration of $T^{\text{ofdm}} = 71.35 , \mu$s. Frequency-shift keying (FSK) was used for ZED modulation, with $F_0 = 125$ Hz and $F_1 = 500$ Hz representing bits 0 and 1, respectively.

To enhance robustness against time jitter and synchronization errors, a 4th-order Butterworth low-pass filter with a cutoff frequency of $f_{cf} = 100$ Hz was applied after the correlation stage (see Figure~\ref{fig:Ccpt_sys}).

Each ZED transmitted a synchronization sequence of duration $T_{\text{seq}} = 800$ ms. To support distinguishable detection, ZED A and ZED B were assigned different periodic transmission cycles: $T_{A} = 1.4$ s and $T_{B} = 2.2$ s, respectively. These offsets ensure that the tags alternate between active and transparent modes, avoiding overlap and enabling separate peak detection.

\begin{figure}[htbp]
\centerline{\includegraphics[scale=0.28]{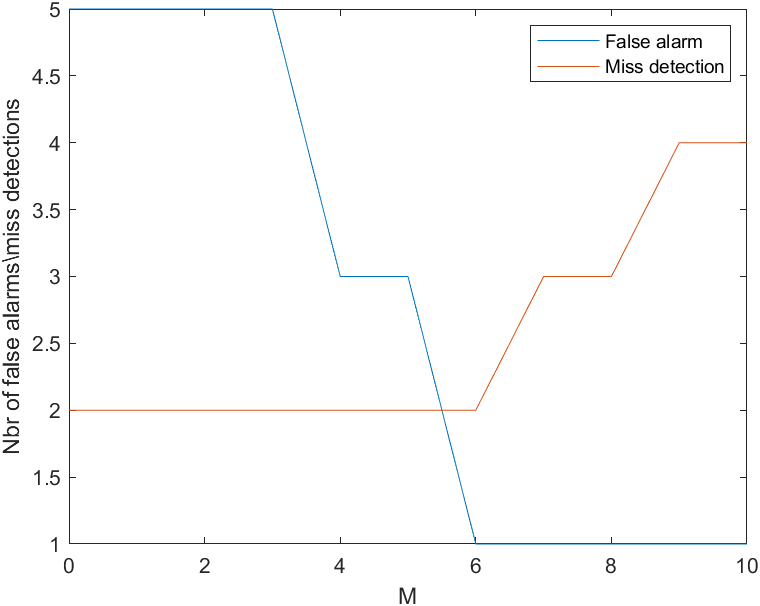}}
\caption{Performance evaluation of secondary peak detection: number of false alarms and missed detections within a 15-second observation window, analyzed across different margin factors M.}
\label{fig:M}
\end{figure}

To distinguish true secondary peaks from side-lobes of a dominant tag, the detection framework employs a secondary threshold defined using a PSL gain margin of $G_{\text{PSL}} = 21.93$ dB and a margin factor $M = 6$ dB to account for noise and interference, as illustrated in figure \ref{fig:M}, we have ferwer false alarms and missed detections when M=6.

\subsection{Performance measurement of the detection system }
\begin{figure*}[htbp]
\centerline{\includegraphics[scale=0.6]{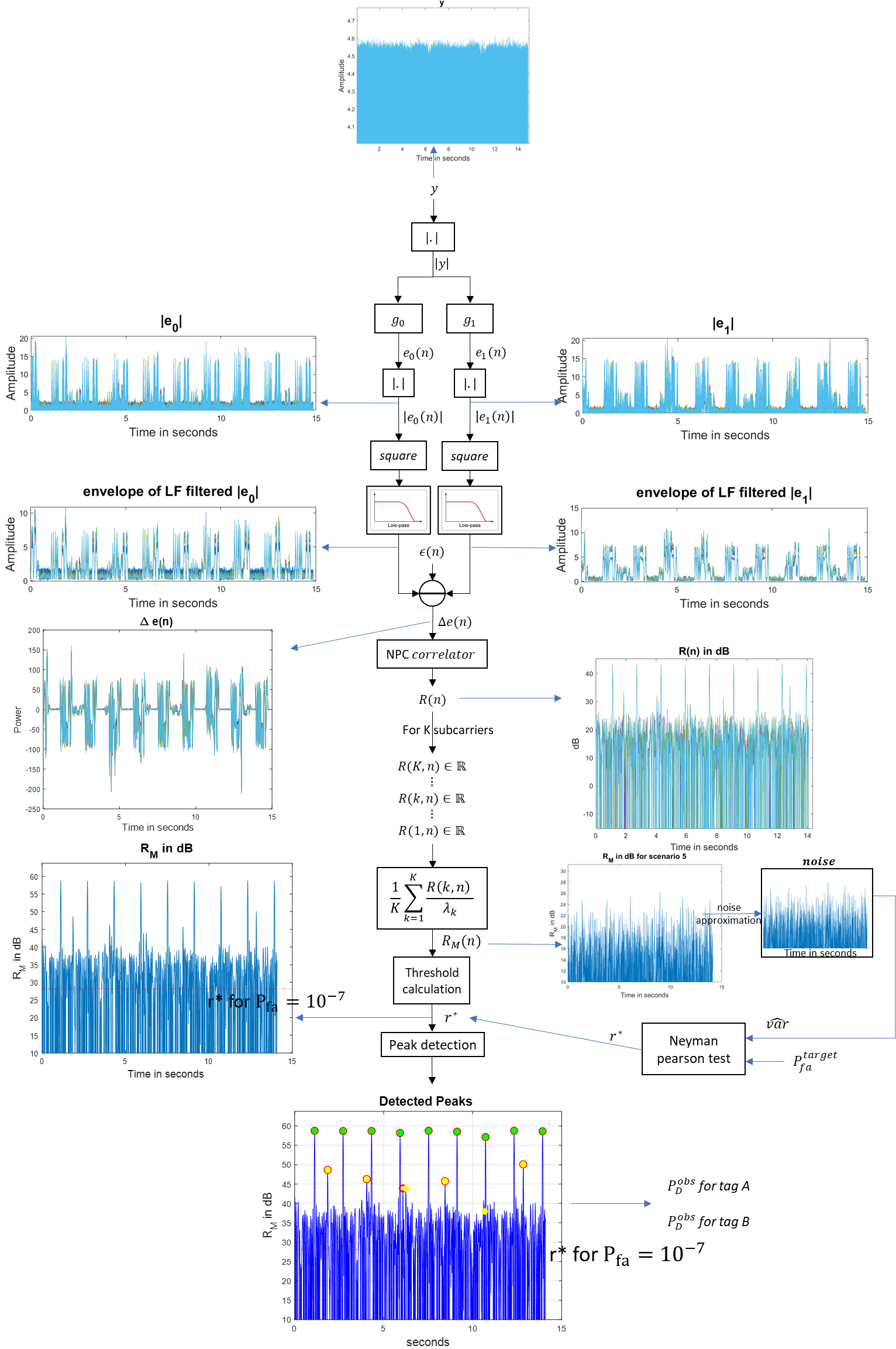}}
\caption{Overview of the experimental detection system and performance evaluation methodology. Includes the signal processing pipeline, detection logic, and observed performance metrics such as false alarm and detection probabilities.}
\label{fig:Perf_sys}
\end{figure*}

\begin{figure*}[htbp]
\centerline{\includegraphics[scale=0.36]{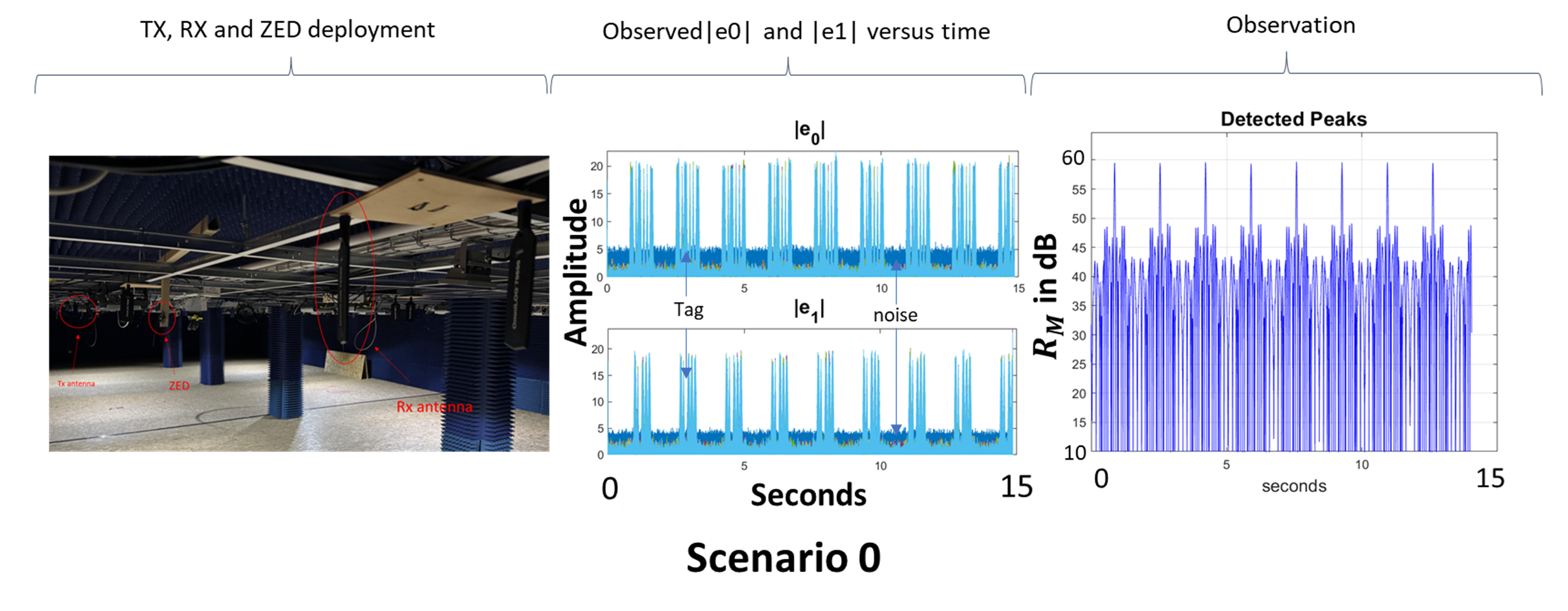}}
\caption{Experimental results for 21-bit Barker code}
\label{fig:scn0}
\end{figure*}

Figure \ref{fig:Perf_sys} presents an overview of the detection system's performance evaluation framework. The central part of the figure, which has been introduced in previous sections, depicts the core signal processing flow for ambient backscatter detection. In this extended representation, the left and right figures illustrate the performance metrics used to assess the system's effectiveness. The last result visualizes the final detection results, including false alarms, missed detections, and successful detections. This expanded view effectively demonstrates the relationship between the detection process and the corresponding performance outcomes.  

\subsection{Experimental results for different scenarios}

\begin{figure*}[htbp]
\centerline{\includegraphics[scale=0.42]{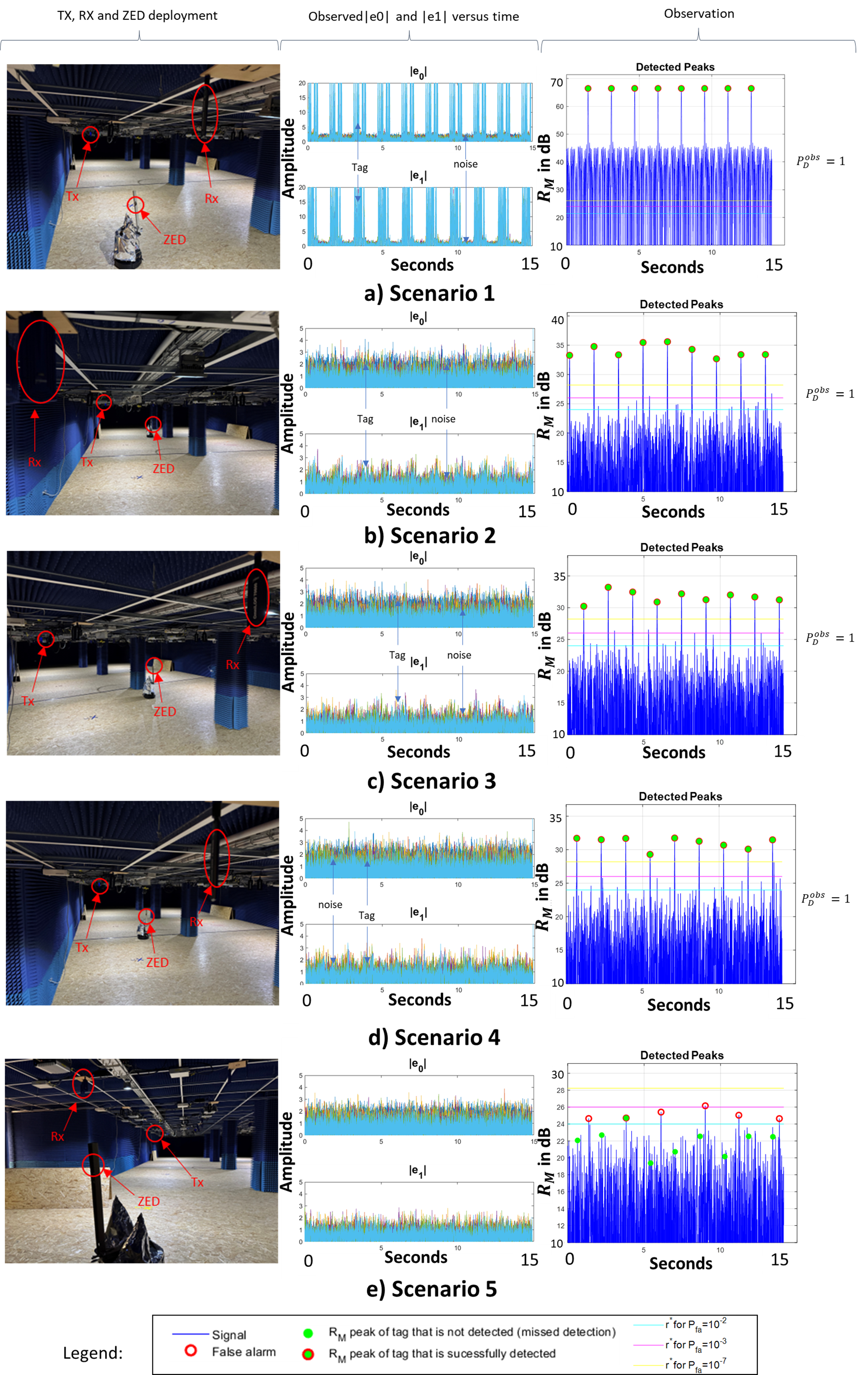}}
\caption{Experimental results for scenarios with a single ZED placed at various positions in the room. Each row shows deployment layout, correlator outputs $(|e_0|,|e_1|)$, and detected contrast peaks. Results demonstrate robust detection performance across different SNR conditions using the proposed NPC.}
\label{fig:monotag}
\end{figure*}

\begin{figure*}[htbp]
\centerline{\includegraphics[scale=0.43]{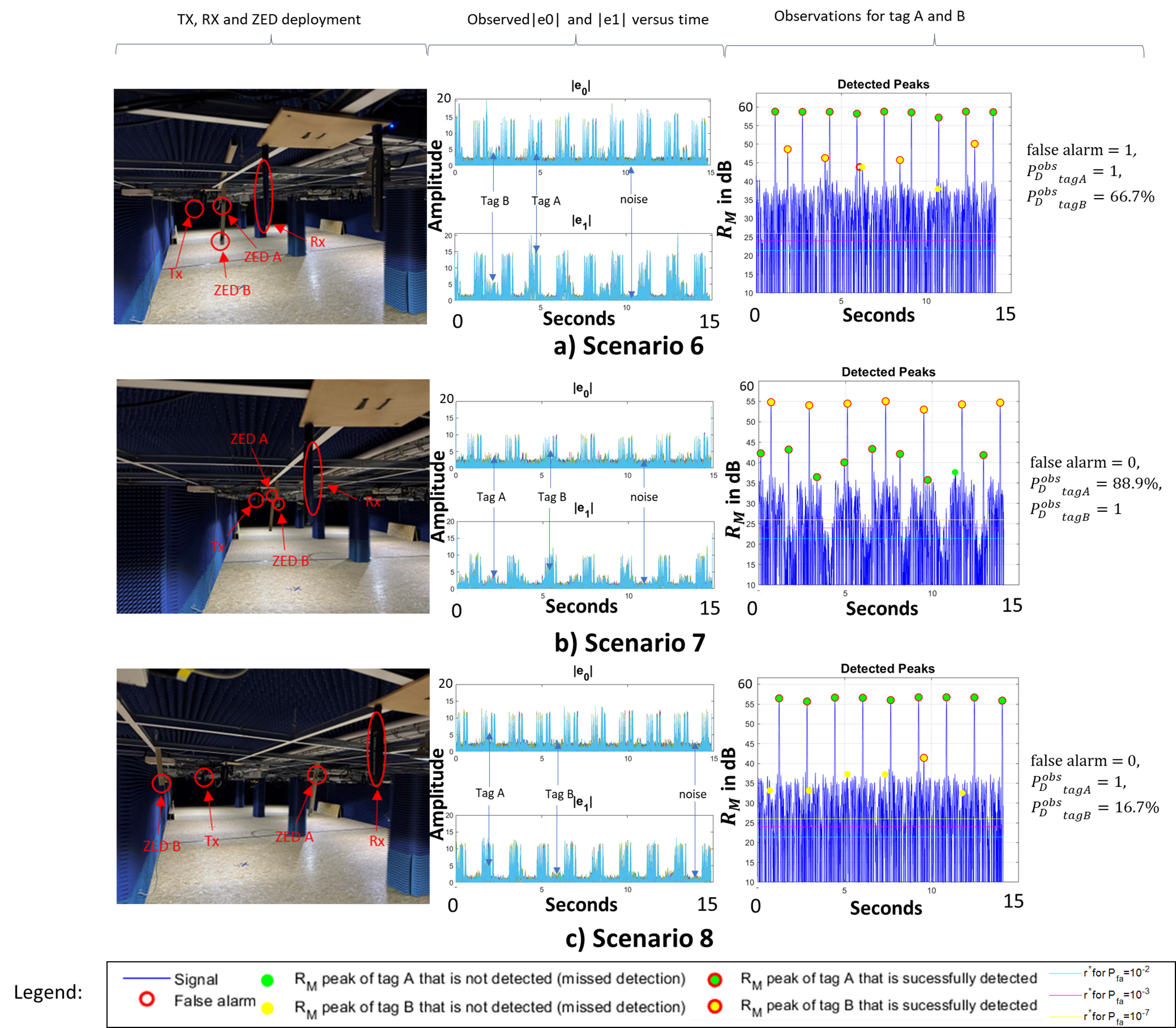}}
\caption{Experimental results for scenarios with two coexisting ZEDs placed at different spatial locations. Detection probability and peak separation are evaluated under various tag configurations, showing strong performance in identifying secondary tags when their signal strength is above the defined threshold.}
\label{fig:multitag}
\end{figure*}

Figure \ref{fig:scn0} presents results from our previous work, which utilized a 21-bit Barker code \cite{yang2025neymanpearsondetectorambientbackscatter}. Figures \ref{fig:monotag}(a) to \ref{fig:monotag}(e) illustrate the results for scenarios 1 to 5, respectively, where a tag is fixed on a robot and placed at different positions within the room, and which utilizes the proposed NPC. Similarly, Figures \ref{fig:multitag}(a) to \ref{fig:multitag}(c) present the results for scenarios 6 to 8, where two tags are positioned differently depending on the specific scenario. For scenario 6, the two tags are placed above and below on a same stick. For scenario 7, they are placed front and behind. And for scenario 8, they are placed left and right on a same stick.  For each scenario, the first column of the figure represents the deployment configuration, including the locations of the transmitter (TX), receiver (RX), and ZED camera. The second column displays the observed values of \( |e_0| \) and \( |e_1| \) over time, while the third column illustrate the observed tag peaks and detection probability \( P_D^{obs} \) for target false alarm probabilities set at \( P_{fa} = 10^{-2}, 10^{-3},10^{-7} \). 

Since the result in scenario 5 is almost full of noise, we can approximate it as gaussian noise and to do the NP test to estimate the thresholds corresponding to target false alarm probabilities. The thresholds are presented as three lines in blue, pink and yellow, used for scenario 1 to 8 in fig.\ref{fig:monotag} and \ref{fig:multitag}.  

In Fig. \ref{fig:scn0}, we provide the detection based on the former sequence used in \cite{yang2025neymanpearsondetectorambientbackscatter}. This figure shows that the PSL is about 10dB, and is not sufficient to permit the detection of a secondary peak. This Figure illustrates the interet of using a high PSL sequence to facilitate the detection of a secondary peak. Roughly we have a high probability to detect a secondary ZED if its power signal is between [-20dB , 0dB] from the signal of the first ZED, which is a significant margin.
By comparing scenario 0 in Fig.\ref{fig:scn0} and scenario 1 in figure \ref{fig:monotag}, we observe that the peak-to-lobe ratio has significantly improved, increasing from approximately 11 dB to 21 dB. This notable enhancement demonstrates the superior performance of our newly proposed NP code, further validating its effectiveness in multiple tags detection. 

From scenarios 1 to 4 presented in figure \ref{fig:monotag}, we observe that our proposed detection scheme using the NPC achieves excellent performance in detecting a single tag. Scenarios 2, 3, and 4 further demonstrate that our model maintains robust detection capabilities even under poor SNR conditions. Specifically, all tag occurrences are successfully detected, with no false alarms or missed detections.

From the results presented in Figure \ref{fig:multitag}, it can be observed that the detection of the first tag consistently exhibits strong significance across all scenarios. Regarding the detection of the second tag within the 15-seconds observation interval, scenario 6 shows that 4 out of 6 peaks are successfully detected, whereas scenario 7 demonstrates even better performance, successfully detecting 8 out of 9 peaks. These results validate that our proposed detection scheme combined with the NP code effectively identifies the second tag under varying conditions. However, in scenario 8, due to the considerably lower detected power level of the second tag, only 1 successful detection is achieved during the 15-second period.

\section{Conclusion}
In this paper, we proposed a robust detection method for multiple Zero-Energy Devices (ZEDs) in ambient backscatter systems using the Neyman-Pearson framework. Building on previous single-tag detection schemes, our approach leverages the Near-Perfect Code (NPC) to improve synchronization and side-lobe suppression. We introduced a new detection strategy that identifies primary and secondary tags based on contrast peaks and defined a side-lobe-aware threshold for secondary ZED detection. The proposed method was validated through experimental scenarios using the CorteXlab testbed, demonstrating reliable detection even under challenging conditions such as low SNR and asynchronous operation. Compared to traditional Barker code-based methods, the NPC provides a significant PSL gain, enabling more accurate multi-tag separation. Future work will focus on tag identification and collision resolution to support a larger number of ZEDs in dynamic environments.

While this work focuses on detection, we acknowledge that collision between tags remains a potential challenge in dense deployments. In particular, the absence of coordination may lead to overlapping transmissions. To address this, the integration of lightweight medium access control (MAC) mechanisms—such as slotted or randomized scheduling inspired by RFID protocols—presents an important direction for future research. Such techniques will be essential for enabling scalable and reliable multi-tag operation in real-world ambient backscatter applications.

\section*{Acknowledgments}
This work is partly supported by the European Project Hexa-X II under (grant 101095759) and by BPI France under the program France Relance (5G Events Labs).



\bibliographystyle{IEEEtran}
\bibliography{references}

\vfill

\end{document}